\RequirePackage{fix-cm}
\documentclass[twocolumn]{svjour3}          

\pdfoutput=1
\usepackage{graphicx}%
\usepackage{multirow}%
\usepackage{amsmath,amssymb,amsfonts}%
\usepackage{mathrsfs}%
\usepackage[title]{appendix}%
\usepackage{xcolor}%
\usepackage{textcomp}%
\usepackage{manyfoot}%
\usepackage{booktabs}%
\usepackage{listings}%
\usepackage[english]{babel}
\usepackage{graphicx}
\usepackage{textcomp}
\usepackage{dsfont}
\usepackage{xcolor}
\usepackage{acronym}
\usepackage{units}
\usepackage{comment}
\usepackage{url}
\usepackage[normalem]{ulem}
\usepackage[round]{natbib}
\usepackage{graphicx}
\usepackage{multicol,caption}

\journalname{Arxiv Preprint}

\acrodef{ICEV}[ICEV]{internal combustion engine vehicle}
\acrodef{GHG}[GHG]{green house gas emission}
\acrodef{BEV}[BEV]{battery electric vehicle}
\acrodef{GWP}[GWP]{global warming potential}
\acrodef{NMC}[NMC]{nickel manganese cobalt}
\acrodef{LCA}[LCA]{life cycle assessment}

\newenvironment{Figure}
{\par\medskip\noindent\minipage{\linewidth}}
{\endminipage\par\medskip}

\newenvironment{Table}
{\par\medskip\noindent\minipage{\linewidth}}
{\endminipage\par\medskip}

\begin{document}
	
	\title{Meta-analysis of Life Cycle Assessments for Li-Ion Batteries Production Emissions %
		\thanks{This publication is part of the project \cite{NEON} (with project number 17628 of the research program Crossover which is (partly) financed by the Dutch Research Council (NWO)).}}
	

%
%
%

\noindent
\author{ \small
		Maurizio Clemente   \and
		Prapti Maharjan		\and
        Mauro Salazar		\and
        Theo~Hofman     }


\institute{M. Clemente, M. Salazar, T. Hofman \at
              Department of Mechanical Engineering, Eindhoven University of Technology \\
              Groene Loper, Eindhoven, 5600MB, North Brabant, Netherlands \\
              \email{maurizio.clement@gmail.com, m.r.u.salazar@tue.nl, t.hofman@tue.nl}           
           \and
           P. Maharjan \at
              Department of Industrial Engineering and Innovation Sciences, Eindhoven University of Technology \\
              Groene Loper, Eindhoven, 5600MB, North Brabant, Netherlands
              \email{p.maharjan@tue.nl}}

\date{Received: date / Accepted: date}

\onecolumn
\maketitle

\begin{abstract}

\textbf{\newline \noindent Purpose} This paper investigates the environmental impact of Li-Ion batteries by quantifying manufacturing-related emissions and analyzing how electricity mix and production scale affect emission intensity.

\noindent\textbf{Methods} To this end, we conduct a meta-analysis of life cycle assessments on lithium-ion batteries published over the past two decades, categorizing them by year, battery chemistry, functional unit, system boundaries, and electricity mix.
We then carry out a cradle-to-gate assessment for a nickel manganese cobalt 811 battery with a silicon-coated graphite anode, analyzing how variations in the carbon intensity of the electricity mix affect emissions, with case studies for China, South Korea, and Sweden.
Finally, we develop a set of regression models that link annual battery production and the carbon intensity of China’s electricity mix to the average mass-specific emissions observed each year.


\noindent\textbf{Results and Discussion} The meta-analysis shows a median global warming potential of 17.63 \unit{kg} CO$_2$-eq./\unit{kg} of battery, with a standard deviation of 7.34. 
Our assessment results closely align: 17.33 \unit{kg} CO$_2$-eq./\unit{kg} for China, 16.85 for South Korea, and 16.47 for Sweden. 
Differences in electricity mix mainly influence emissions from the energy-intensive cell production, particularly from cathode material processing.
We used the data gathered in the meta-analysis to analyze and compare several regression models. We found that a multivariate linear regression using production volume and the carbon intensity of the Chinese electricity mix as predictors explains emissions with moderate accuracy (R$^2$ = 0.6034).


\noindent\textbf{Conclusion} The environmental impact of battery manufacturing can be reduced by using clean energy sources in production processes.
However, achieving substantial reductions requires clean energy throughout the entire supply chain, as importing materials from regions with carbon-intensive electricity mixes can undermine these efforts.
Our findings also highlight the emission-reducing effect of learning associated with increased production scale, supporting the integration of learning effects in future life cycle assessment models.


\keywords{Lithium-Ion Batteries \and Life Cycle Assessment \and Meta-analysis \and Environmental Impact}

\end{abstract}

\vspace{40pt}

\begin{multicols}{2}

\begin{Figure}
	\centering
	\includegraphics[width=0.9\linewidth,keepaspectratio]{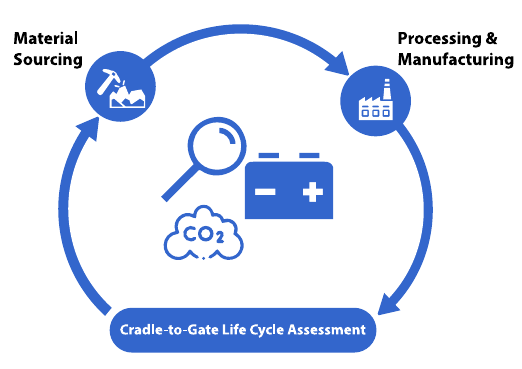}
	\captionof{figure}{In our study, we analyze the emissions generated during battery production according to several studies from the last two decades, we carry out a cradle-to-gate life-cycle assessment accounting for the contribution of the carbon intensity of the electricity mix, and we correlate the average mass-specific emissions to the annual production volumes and the carbon intensity of the Chinese electricity mix.}
	\label{fig:Header1}
\end{Figure}

\section{Introduction}~\label{sec:Introduction}
Global \acp{GHG} caused by human activities have been unequivocally held responsible for global warming~\citep{AR6}.
In the collective effort to reduce environmental impact, many researchers have focused on the transportation sector, primarily targeting light-duty vehicles.
Energy-efficient \acp{BEV} have been proposed as a solution to mitigate the environmental impact of road transport without severely compromising the modern paradigm of mobility and are rapidly gaining market shares.
While the lack of tailpipe emissions and higher powertrain efficiency contribute to reducing local pollution compared to conventional internal combustion engines, \acp{BEV} still require electric energy to operate.
If this energy comes from non-renewable sources, there is a risk of shifting the emissions upstream in the energy supply chain, rather than addressing the problem directly. 
Even when using renewable energy sources, such as photovoltaic plants and wind farms, there is an increasing demand for energy storage to power vehicles, accommodate power peaks in production facilities, and prevent grid overloading.
Nowadays, batteries have become an essential component of modern systems, providing energy storage for a wide variety of applications.
However, battery production (Fig.~\ref{fig:Header1}) involves energy-intensive manufacturing processes and rare-earth materials~\citep{AichbergerJungmeier2020,ChordiaNordeloefEtAl2021}, which are frequently overlooked sources of \acp{GHG} emissions.
Global \acp{GHG} emissions caused by light-duty vehicles operations, accounting for 10.4\% of the total~\citep{IEA2020}, could end up simply being reclassified under other sectors, such as industry and power generation, jeopardizing our efforts to limit \acp{GHG}.
Although the overall contribution of battery production to global emissions is expected to increase~\citep{RitchieRoserEtAl2020,Rawles2019}, it is crucial to estimate the emissions generated during battery manufacturing to gain a clear understanding of the extent to which we are actually reducing emissions. 
In this context, the projected increase in battery demand has led to the development of facilities with larger production capacities (giga-factories)~\citep{Rawles2019,YuSumangil2021}, which leverage learning effects to reduce costs and emissions. 
Therefore, in order to accurately forecast the emissions related to battery manufacturing, it is crucial to take into account the impact of learning in the production process.
Against this backdrop, our paper examines the literature on the environmental impacts of batteries, presents a cradle-to-gate \ac{LCA} study for estimating the manufacturing-related emissions of rechargeable, prismatic, \ac{NMC} 811 Li-Ion batteries with silicon-coated graphite-based anode, and proposes regression models to predict trends in specific emissions.




\emph{Related Literature:}
This paper pertains to the \ac{LCA} research line and inquires about learning effects owing to the production volumes.
As outlined in ISO 14040~\citep{ISO2020,ISO2006,ISO2006b}, \ac{LCA} provides a structured approach to evaluate the total impact of a product or system, such as environmental or economic effects, across all stages of its life, from raw material extraction through manufacturing, distribution, use, and final disposal.
An \ac{LCA} generally involves four main stages: defining the goal and scope, compiling the life-cycle inventory, conducting the impact assessment, and interpreting the findings.
The goal and scope phase clarifies the objective of the assessment and specifies the functional unit used to measure the product’s impact~\citep{ArzoumanidisDEusanioEtAl2020}.
The inventory phase captures all relevant inputs and outputs, such as materials, energy, emissions, and waste, within the defined system boundaries.
The impact assessment then translates this data into indicators aligned with the functional unit.
Lastly, the interpretation phase links the outcomes back to the original objectives and supports insightful conclusions.
In one of the first papers on the topic~\citep{NotterGauchEtAl2010}, the authors investigate the overall mass-specific environmental impact of lithium-ion batteries used in \acp{BEV} compared to \acp{ICEV}.
Specifically, they address the role of production, operation, and disposal, concluding that despite the additional burden from battery production, \acp{BEV} are environmentally advantageous over ICEVs.
A considerable number of subsequent studies focused on different aspects, such as the impact of solvents used in production~\citep{ZackrissonAvellanEtAl2010}, recycling~\citep{CusenzaBobbaEtAl2019}, depth-of-discharge~\citep{Majeau-BettezHawkinsEtAl2011,RydhSanden2005}, and battery lifespan~\citep{AmarakoonSmithEtAl2012}.
Overall, most studies agree that materials and energy consumption in industrial processes are major contributors to emissions, largely due to the dependence on the carbon intensity of the electricity mix~\citep{DunnGainesEtAl2014}, while material transportation plays a minimal role~\citep{RydhSanden2005}.
However, retrieving an unanimous estimation of the life-cycle emissions proves very difficult due to the lack of harmonization in functional units, system boundaries, battery chemistry, and regional parameters.
In fact, despite the ISO14040 standard, the purpose and focus of studies in the literature vary widely.
Some authors measure emissions per kilometer driven~\citep{ZackrissonAvellanEtAl2010, GiordanoFischbeckEtAl2018, LaPicirellideSouzaSilvaLoraEtAl2018,DengMaEtAl2019}, others per kilogram of battery~\citep{YinHuEtAl2019,AguirreEisenhardtEtAl2012,SullivanGaines2012}, and others per kWh of battery capacity~\citep{QiaoZhaoEtAl2017,KallitsisKorreEtAl2020, SunLuoEtAl2020,DaiKellyEtAl2019}.
Additionally, they all examine different vehicles, with varying driving ranges and batteries of different technologies and sizes.
The diversity of this set reflects the complexity of assessing environmental impacts in battery production, with energy sources, databases, material choices, and manufacturing methods all playing essential roles in total emissions.
Some recent works~\citep{RomareDahlloef2017,ZhaoWalkerEtAl2021} provided a comparison of different studies to give a picture of how and why results differ, and what the most probable figures are, underlining the importance of standardized units for future assessments.
Although standard \ac{LCA} does not consider the learning effects, integrating these influences in manufacturing processes to account for past and future scenarios provides more realistic perspectives on emissions~\citep{SohnKalbarEtAl2019}.
The idea behind process learning first appeared in the academic world when Theodore Wright discovered that the cost of producing aircraft declined as a function of cumulative production~\citep{Wright1936,Kahouli-Brahmi2008}.
This effect has been called ``learn-by-doing" and originates from the experience gained in repeating a process, such as increased labor productivity or efficiency in the use of raw materials.
Previous work that delved into learning effects found that up-scaling facilities reduces energy demand during production, improves material efficiency, and lowers environmental impacts~\citep{DavidssonKurland2020,BergesenSuh2016}.
Although numerous authors have conducted \acp{LCA} of the battery supply chain~\citep{ChordiaNordeloefEtAl2021,RomareDahlloef2017,ZhaoWalkerEtAl2021}, to the best of the authors' knowledge, a thorough meta-analysis of the battery manufacturing emissions using a statistical approach, and the influence of learning effects owing to the annual battery production and the electricity mix have been overlooked.

\emph{Statement of Contribution:}
In this study, we conduct an exhaustive meta-analysis of \ac{LCA} studies examining the environmental impact of batteries over the past two decades. 
We categorize each paper by publication year, battery dimension, chemistry, functional unit, major contribution, system boundaries, and electricity mix.
Next, we perform a cradle-to-gate \ac{LCA}, based on \citep{DaiDunnEtAl2017,DaiKellyEtAl2018,DaiKellyEtAl2019}, to assess the emissions generated during the manufacturing of one kilogram of a rechargeable, prismatic, \ac{NMC} 811 Li-Ion battery with silicon-coated graphite-based anode, presenting a case study considering the impact of the electricity mix in countries with varying carbon intensities: Sweden, South Korea, and China.
We propose different regression models to estimate the specific manufacturing-related battery emissions based on the annual production and the carbon intensity of the Chinese electricity mix, enabling the estimation of (future) \acp{GHG}.
Ultimately, energy transition researchers and engineers could employ these models to estimate emissions and achieve a footprint-aware design~\citep{Clemente2025PhD}, which has potential applications in various domains.



\emph{Organization:}
The remainder of this study is structured as follows: Section~\ref{sec:Methodology2} presents a review of \ac{LCA} studies on the environmental impacts of batteries, along with our cradle-to-gate \ac{LCA} framework and the proposed regressions.
Section~\ref{sec:Results2} applies our \ac{LCA} methodology in a case study, presents the regression analysis results, and discusses the main findings.
The conclusions of the study are summarized in Section~\ref{sec:Conclusions}, along with recommendations for future research.


\section{Method}~\label{sec:Methodology2}

In this section, we illustrate the methodology for GHG estimation in detail.
Section~\ref{subsec:Meta} presents a meta-analysis of the literature, examining factors that influence emissions generated during battery production and consolidating two decades of research into a dataset.
Section~\ref{subsec:LCA} outlines our \ac{LCA} methodology, detailing the manufacturing process and defining the system boundaries.
Section~\ref{subsec:Reg} introduces regression models to predict the specific environmental impact of batteries, and distill insights from these regressions into a quantitative learning effect.
Finally, Section~\ref{subsec:disc} discusses the assumptions and limitations of our approach.

\subsection{Meta-analysis}~\label{subsec:Meta}

\begin{table*}
	\centering
	\caption{Summary of Li-Ion Battery \ac{LCA} studies. The global warming potential (GWP) is expressed in \unit{kg} CO$_2$-eq and MCF stands for most contributing factor.\\}
	\renewcommand{\arraystretch}{1.2} 
	\resizebox{\textwidth}{!}{
		\begin{tabular}{l c c c c c c c c}
			\hline
			\textbf{Author(s)} & \textbf{Year} & \textbf{Battery Chemistry} & \textbf{Dimensions} & \textbf{Functional Unit} & \textbf{\begin{small} Most Contributing Factor
			\end{small} } & \textbf{System Boundaries} & \textbf{Electricity Mix} & \textbf{GWP} \\
			\hline
			
			\cite{DunnGainesEtAl2014} & 2011 & NMC-C & 28 kWh & Mass (g) & Energy (Manufacturing) & Cradle-to-gate & USA, California & 22 \\
			
			\cite{EllingsenSinghEtAl2013} & 2013 & NMC-C & 12 kWh & Mass (kg), Capacity (kWh) & Energy (Cell Manufacturing) & Cradle-to-gate & South Korea, Norway & 18.12 \\
			
			\cite{HawkinsSinghEtAl2013} & 2013 & NMC-C & 24 kWh & Distance (km) & Energy (Use-phase) & Cradle-to-grave &  European avg. & 21.59 \\
			
			\cite{LiGaoEtAl2014} & 2014 & NMC-SiNW & 43.2 kWh & Distance (km) & Energy (Use-phase) & Cradle-to-grave & China & 96.96 \\
			
			\cite{EllingsenSinghEtAl2016} & 2016 & NMC-C & 17.7 kWh & Distance (km) & Energy (Use-phase) & Cradle-to-grave & Norway & 15.82 \\
			\cite{EllingsenSinghEtAl2016} & 2016 & NMC-C & 24.4 kWh & Distance (km) & Energy (Use-phase) & Cradle-to-grave & Norway & 15.42 \\
			\cite{EllingsenSinghEtAl2016} & 2016 & NMC-C & 42.1 kWh & Distance (km) & Energy (Use-phase) & Cradle-to-grave & Norway & 14.5 \\
			\cite{EllingsenSinghEtAl2016} & 2016 & NMC-C & 59.9 kWh & Distance (km) & Energy (Use-phase) & Cradle-to-grave & Norway & 14.29 \\
			
			\cite{KimWallingtonEtAl2016} & 2016 & LMO/NMC-C & 24 kWh & Mass (kg), Capacity (kWh) & Energy (Cell Manufacturing) & Cradle-to-gate & South Korea & 11.1 \\
			
			\cite{LuWuEtAl2016} & 2016 & NMC-C & 20 kWh & Capacity (kWh) & Materials (Rare Earths) & Cradle-to-gate & China & 7.29 \\
			
			\cite{WangYuEtAl2016} & 2016 & NMC-C & 16 kWh & Capacity (kWh) & Energy (Use-phase) & Cradle-to-grave & China & 68.17 \\
			\cite{WangYuEtAl2016} & 2016 & L(R)NMO-C & 16 kWh & Capacity (kWh) & Energy (Use-phase) & Cradle-to-grave & China & 79.81 \\
			
			\cite{Zackrisson2016} & 2016 & NMC-Li & 5 Ah cell & Distance (km) & Energy (Use-phase) & Cradle-to-grave & EU and Global avg. & 23.04 \\
			
			\cite{DengLiEtAl2017} & 2017 & NMC-C & 63.8 kWh & Distance (km) & Energy (Use-phase) & Cradle-to-grave & USA & 21.65 \\
			\cite{DengLiEtAl2017} & 2017 & NMC-MoS\textsubscript{2} & 49.4 kWh & Distance (km) & Energy (Use-phase) & Cradle-to-grave & USA & 29.53 \\
			\cite{DengLiEtAl2017} & 2017 & NMC-C & 66 kWh & Distance (km) & Energy (Use-phase) & Cradle-to-grave & USA & 19.01 \\
			
			\cite{HaoMuEtAl2017} & 2017 & NMC-C & 28 kWh & Capacity (kWh) & Materials (Cathode) & Cradle-to-gate & China & 17.13 \\ %
			
			\cite{QiaoZhaoEtAl2017} & 2017 & NMC-C & 170 kg & Capacity (kWh) & Materials (Steel for Body) & Cradle-to-gate & China & 16.94 \\ %
			\cite{QiaoZhaoEtAl2017} & 2017 & NMC-C & 170 kg & Capacity (kWh) & Materials (Steel for Body) & Cradle-to-gate & China & 15.91 \\%
			
			
			\cite{DaiKellyEtAl2018} & 2018 & NMC111 & 23.5 kWh & Capacity (kWh) & Materials (Cathode) & Cradle-to-gate & USA & 16.11 \\ %
					
			\cite{GiordanoFischbeckEtAl2018} & 2018 & NMC333-C & 70.2 kWh & Distance (km) & Energy (Use-phase) & Cradle-to-grave & Norway & 26.25 \\%
			\cite{GiordanoFischbeckEtAl2018} & 2018 & NMC333-C & 46.8 kWh & Distance (km) & Energy (Use-phase) & Cradle-to-grave & Norway & 13.13 \\%
			\cite{GiordanoFischbeckEtAl2018} & 2018 & NMC333-C & 23.4 kWh & Distance (km) & Energy (Use-phase) & Cradle-to-grave & Norway & 8.75 \\%
			\cite{GiordanoFischbeckEtAl2018} & 2018 & NMC441-C & 70.2 kWh & Distance (km) & Energy (Use-phase) & Cradle-to-grave & Norway & 25.4 \\%
			\cite{GiordanoFischbeckEtAl2018} & 2018 & NMC441-C & 46.8 kWh & Distance (km) & Energy (Use-phase) & Cradle-to-grave & Norway & 12.7 \\%
			\cite{GiordanoFischbeckEtAl2018} & 2018 & NMC441-C & 23.4 kWh & Distance (km) & Energy (Use-phase) & Cradle-to-grave & Norway & 8.47 \\%
			
			\cite{WuKong2018} & 2018 & NMC-C & 100 kWh & Mass (kg), Capacity (kWh) & Energy (Cell Manufacture) & Cradle-to-gate & EU & 21.35 \\%
			\cite{WuKong2018} & 2018 & NMC-Li & 100 kWh & Mass (kg), Capacity (kWh) & Energy (Cell Manufacture) & Cradle-to-gate & EU & 23.65 \\%
			\cite{WuKong2018} & 2018 & NMC-SiNWs & 100 kWh & Mass (kg), Capacity (kWh) & Energy (Cell Manufacture) & Cradle-to-gate & EU & 36.33 \\ %
			
			
			\cite{YuWeiEtAl2018} & 2018 & NMC-C & 123 kg & Distance (km) & Energy (Use-phase) & Cradle-to-grave & China & 57.39 \\%
			
			\cite{CusenzaBobbaEtAl2019} & 2019 & LMO-NMC & 11.4 kWh & Capacity (kWh) & Energy (Manufacturing) & Cradle-to-grave & Japan & 25.83 \\%
			
			\cite{DaiKellyEtAl2019} & 2019 & NMC111-C & 23.5 kWh & Capacity (kWh) & Materials (Cathode) & Cradle-to-gate & USA & 10.38 \\%
			
			\cite{YinHuEtAl2019} & 2019 & NMC333-C & 1 kWh & Mass (kg) & Materials (Cathode) & Cradle-to-gate & USA & 13.92 \\%
			\cite{YinHuEtAl2019} & 2019 & NMC333-C & 1 kWh & Mass (kg) & Materials (Cathode) & Cradle-to-gate & China & 26.4 \\%
											
			\cite{KallitsisKorreEtAl2020} & 2020 & NMC333-C & 26.6 kWh & Capacity (kWh) & Energy (Cell Manufacturing) & Cradle-to-gate & China & 27.59 \\%
			\cite{KallitsisKorreEtAl2020} & 2020 & NMC333-SiC & 40.9 kWh & Capacity (kWh) & Energy (Cell Manufacturing) & Cradle-to-gate & China & 29.05 \\%
			\cite{KallitsisKorreEtAl2020} & 2020 & NMC622-SiC & 46.2 kWh & Capacity (kWh) & Energy (Cell Manufacturing) & Cradle-to-gate & China & 29.25 \\%
			\cite{KallitsisKorreEtAl2020} & 2020 & NMC811-SiC & 52.9 kWh & Capacity (kWh) & Energy (Cell Manufacturing) & Cradle-to-gate & China & 29.33 \\%
			
			\cite{KellyDaiEtAl2020} & 2020 & NMC111-C & 27 kWh & Capacity (kWh) & Materials (Cathode) & Cradle-to-gate & USA, China, SK, JP, EU & 10.59 \\%
			
			\cite{SunLuoEtAl2020} & 2020 & NMC622-C & 72.5 kWh & Capacity (kWh) & Materials (Cathode) & Cradle-to-grave & China & 10.49 \\%
			\hline
		\end{tabular}
	}

	\label{tab:LR}
\end{table*}

\begin{table*}[!htp]
	\centering
	\label{tab:stats}
	\captionof{table}{Statistical analysis of the dataset in Table~\ref{tab:LR}. $N$ represents the dimension of the dataset.\\}
	\resizebox{\textwidth}{!}{
		\begin{tabular}{l | c c c c c c | c c c c c c}
			\hline
			\multirow{2}{*}{\textbf{Dataset}} & \multicolumn{6}{c}{\textbf{With Outliers}} & \multicolumn{6}{c}{\textbf{Without Outlier}} \\
			& N & Average & Median & Std. Dev. & Variance & Range & N & Average & Median & Std. Dev. & Variance & Range \\
			\hline
			\textbf{Whole Dataset} & 40 & 24.77 & 20.18 & 19.09 & 364.39 & 89.67 & 36 & 19.12 & 17.63 & 7.34 &53.80 &29.04\\
			\textbf{Distance [\unit{km}]} & 17 & 24.93 & 19.01 & 21.72 & 471.42 & 88.49 & 15 & 17.97 & 15.82 & 6.40 & 41.00 & 21.06 \\
			\textbf{Capacity [\unit{kWh}]} & 15 & 26.26 & 17.13 & 22.97 & 527.72 & 72.52 & 13 & 18.91 & 16.94 & 7.21 & 52.00 & 22.04 \\
			\textbf{Mass [\unit{kg}]} & 8 & 21.61 & 21.68 & 7.93 & 62.87 & 25.23 & 8 & 21.61 & 21.68 & 7.93 & 62.87 & 25.33 \\
			\hline
		\end{tabular}
	}
\end{table*}

We examined the literature from the last decades of research on battery manufacturing-related emissions, categorizing each study by year, battery chemistry considered, battery dimensions analyzed, functional unit adopted, system boundaries, and electricity mix.
Thereafter, we analyzed their results to identify the primary contributing factor to the battery's environmental impact. 
Among the various environmental impact categories defined in ISO 14040~\citep{ISOFramework}, we focused on global warming potential (GWP), which quantifies greenhouse gas emissions that contribute to climate change.
The standard unit used to measure GWP is \unit{kg CO$_2$}-equivalents.
Since different gases contribute to climate change at varying levels, this unit allows for a common comparison by converting their effects into the amount of \unit{CO$_2$} that would cause the same amount of warming over a specified period.
By using \unit{kg CO$_2$}-equivalents, emissions from different greenhouse gases can be aggregated into a single metric, making it easier to assess and compare the overall climate impact of various processes, including battery manufacturing.
Our analysis does not consider other environmental impact categories (e.g., Ozone Depletion Potential, Acidification Potential), focusing exclusively on GWP to develop a statistical model of the emissions generated during battery production.

\begin{Figure}
	\centering
	\includegraphics[width=0.9\columnwidth]{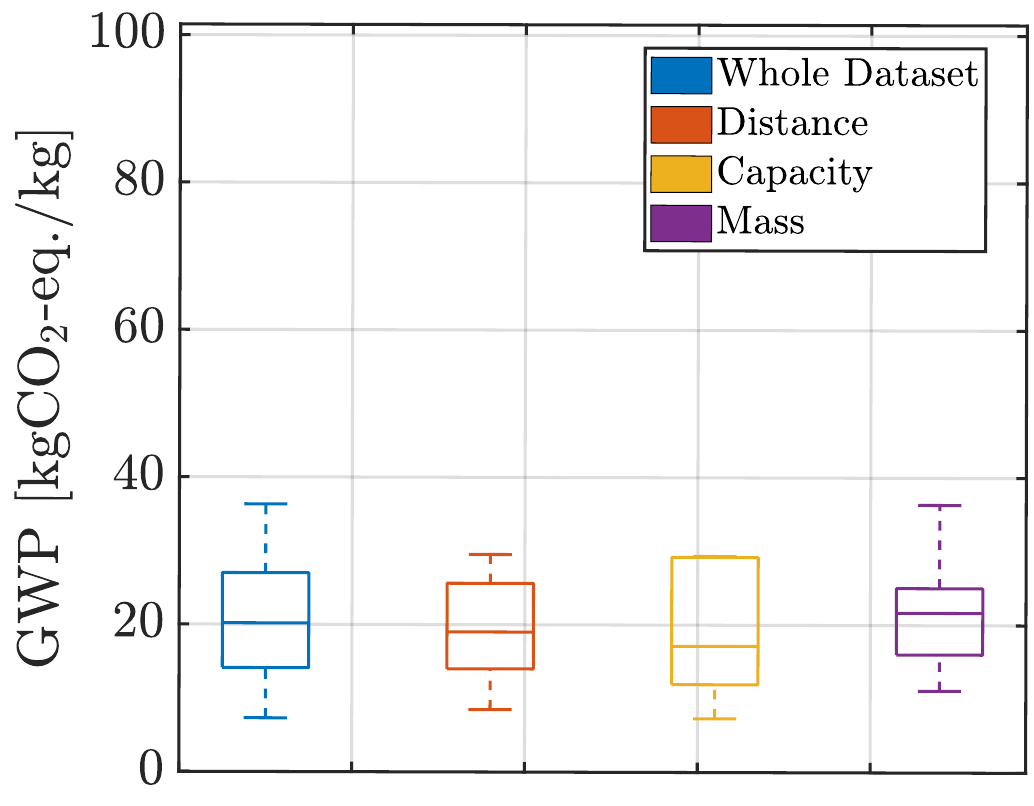}
	\captionof{figure}{Comparison of the average specific environmental impact per functional unit in the dataset (converted to \unit{kg} CO$_2$-eq. per \unit{kg}). The global warming potential is abbreviated as GWP.}
	\label{fig:stats}
\end{Figure}

We condense the findings of the meta-analysis in Table~\ref{tab:LR}.
Leveraging the insights gathered in the analysis, we validate the result of our \ac{LCA} from Section~\ref{subsec:LCA} and assemble a dataset that we can use to develop regression models in Section~\ref{subsec:Reg}.
According to the data, the average value for the mass-specific environmental impact of \ac{NMC} Lithium Ion batteries is 24.77 \unit{kg} CO$_2$-eq. per \unit{kg}, with a median of 20.18 \unit{kg} CO$_2$-eq. per \unit{kg}.
However, the gap between these two values suggests the presence of outliers that elevate the average, hence, the median is a more reliable measure to use.
These outliers correspond to cradle-to-grave assessments conducted in China~\citep{LiGaoEtAl2014,WangYuEtAl2016,YuWeiEtAl2018}, where the significant impact is primarily attributed to energy consumption during the use phase, due to the high carbon intensity of the electricity mix~\citep{HaoMuEtAl2017}.
After excluding outliers, the mean decreases to 19.12 \unit{kg} CO$_2$-eq. per \unit{kg}, and the median to 17.63, with the standard deviation significantly reducing from 19.09 to 7.34 \unit{kg} CO$_2$-eq. per \unit{kg}.
Thereafter, we separately examine the studies in the dataset based on the functional unit adopted and provide a graphical representation of the different descriptive statistics in Fig.~\ref{fig:stats}. 
Each unit fully reflects the purpose of the study and the background of the authors.
For instance, authors with a background in automotive industry often make use of the distance driven \unit{(km)} as unit.
However, the distance driven still depends on the weight of the vehicle used, its lifetime, and could present ambiguities regarding the electricity mix of country of production or usage~\citep{HawkinsSinghEtAl2013}.
For this reason, in our assessment we consider the mass in \unit{kg} as the functional unit, whilst providing a conversion in \unit{kWh}.
Even though it is still possible to convert the units and estimate the emissions of a study in a different functional unit, direct comparisons between units are not always possible, as they assume a linear relationship in emissions scaling with respect to those variables, which is not universally applicable.
Therefore, a standardized functional unit tailored to the specific use case or system boundaries could facilitate comparisons across studies by harmonizing the data.

The main factor contributing to \acp{GHG} generated during battery production depends heavily on the system boundaries and on the electricity mix in the country of production and usage~\citep{NotterGauchEtAl2010,TagliaferriEvangelistiEtAl2016}.
For most cradle-to-grave assessments, the energy consumption during the battery use-life leads to the largest share of emissions.
In this case, the electricity mix in the country of use becomes very impactful in estimating the emissions, suggesting that reducing the carbon intensity of the electricity mix of the county where the \acp{BEV} are used would significantly decrease the overall emissions during the battery life-cycle~\citep{FariaMarquesEtAl2014,EllingsenSinghEtAl2016}.
Concerning the cradle-to-gate assessments, we find different factors depending on the electricity mix in the country of production~\citep{DunnGainesEtAl2014}.
Studies set in geographical locations with prevalence of nuclear or renewable energies (e.g., Norway~\citep{EllingsenSinghEtAl2016} and Sweden~\citep{ChordiaNordeloefEtAl2021}) report reduced overall emissions and identify material extraction as the largest contributor.
Conversely, studies based on regions with higher fossil fuel reliance (e.g., China~\citep{QiaoZhaoEtAl2017,YinHuEtAl2019,LuWuEtAl2016}, Brazil~\citep{LaPicirellideSouzaSilvaLoraEtAl2018}) point to manufacturing energy~\citep{ZackrissonAvellanEtAl2010}.
In general, many studies highlight cathode production, battery management systems, and structural materials like aluminum and copper as significant contributions due to material extraction and energy intense processing~\citep{EllingsenSinghEtAl2016,NotterGauchEtAl2010,HaoMuEtAl2017}.
A few investigations include the impacts of recycling processes, second-life uses in stationary applications, and capacity loss over time~\citep{FariaMarquesEtAl2014}.
Although recycling can reduce the environmental impact, the extent of its contribution largely varies across studies~\citep{DunnGainesEtAl2014}.
Finally, in all the studies, transportation has minimal impact.

\subsection{Life-cycle Assessment}\label{subsec:LCA}

In this study, we devise an \ac{LCA} to estimate the weight-specific \ac{GWP} caused during battery manufacturing and identify the specific processes that have the greatest impact on \acp{GHG}.
Out of the several existent technologies, we focus our attention on a Lithium Ion \ac{NMC} 811 cathode battery chemistry with silicon-coated graphite-based anode, as it is the most common type used in \acp{BEV} with 71\% of the total vehicles sold and is expected to drive the soaring battery production~\citep{IEA2021}.
We consider a cradle-to-gate \ac{LCA} approach, analyzing the Global Warming Potential (GWP) expressed in equivalent kilograms of $\mathrm{CO_2}$  produced per kilogram of battery.
Even though the scientific community is still divided about the choice of the functional unit, in our study, we refer all the flows to the manufacturing of one kilogram of battery, as this approach offers a more straightforward physical interpretation of the flows in relation to the quantities of materials required for the manufacturing processes.
We detail the LCA process to compute the final GWP scores in Fig.~\ref{fig:LCA1}.
We carry out the inventory phase by defining the system and its functional unit (1 \unit{kg} of battery) and considering all the processes that lead to the final battery assembly from the raw material extraction (cradle-to-gate analysis).
To this end, we use the inventory described in the GREET model~\citep{DaiDunnEtAl2017,DaiKellyEtAl2018,DaiKellyEtAl2019} to consider all the flows of materials, energy, waste products, and emissions in the system.
Once every flow is identified, we retrieve the data on the associated emission from the Ecoinvent 3.8 database~\citep{FrischknechtJungbluthEtAl2005,WernetBauerEtAl2016}.
We use the software ``Activity Browser"~\citep{SteubingKoningEtAl2020} to efficiently change parameters, providing significant flexibility in running alternative scenarios, such as switching to different electricity mixes for each process.
Thereafter, we adopt the ``ReCipe Midpoint Hierarchist method" to translate inventory results (e.g., emissions and resource use) into environmental impact scores.
Midpoint indicators are problem-oriented~\citep{HuijbregtSteinmannEtAl2017}, focusing on a single environmental problem, such as climate change or acidification, while endpoint indicators show the environmental impact on higher aggregation levels, such as ecosystem damage or human health.
In this study, we only consider the midpoint indicators for the climate change category, leaving the others (photo-chemical ozone formation, terrestrial acidification, freshwater eutrophication, land use and fossil resource scarcity) to potential extensions of the framework in the future, allowing for potential trade-offs.
Finally, we leverage the results of the analysis to evaluate how and how much the different processes, materials and energy sources influence the battery manufacturing emissions.

\subsection{Regression Analysis}~\label{subsec:Reg}

In recent years, a growing body of literature has examined how the production costs of batteries will evolve over time~\citep{MaulerDuffnerEtAl2021}, as technology advances.
While the scale of production has been correlated with an expected reduction in costs~\citep{Hoekstra2019,ClementeSalazarEtAl2025}, its effect on the emissions has been studied only to a limited extent~\citep{PhilippotAlvarezEtAl2019}.
In our study, we aim to investigate the learning effects taking place for the environmental impact of lithium batteries by proposing different regression models, based on the annual battery production (in \unit{GWh}) $Q_\mathrm{a}$ and the carbon intensity of the Chinese electricity mix (in \unit{g} CO$_2$/\unit{kWh}) $E_\mathrm{ch}$, shown in Figures \ref{fig:trend_AN} and \ref{fig:trend_CH}.
To this end, we analyze the evolution over time of specific emissions $P_{\mathrm{gw}}$ and the yearly-averaged specific emissions $\overline{P}_{\mathrm{gw}}$, using a linear and power-law functions to fit the data.
Hereby, we present the regressions used for the former, while the corresponding equations for the latter can be derived analogously.
First, we leverage linear regressions in the form
\begin{equation}\label{eq:GWP-Q-l}
	P_{\mathrm{gw}}  = \beta_{\mathrm{l},0} + \beta_{Q_\mathrm{a,l},1} \left(Q_\mathrm{a}\right),
\end{equation}
\begin{equation}
	P_{\mathrm{gw}}  = \beta_{\mathrm{l},0} + \beta_{E_\mathrm{ch,l},1} \left(E_\mathrm{ch}\right),
\end{equation}
and both terms at the same time in a multi-variate regression
\begin{equation}
	P_{\mathrm{gw}}  = \beta_{\mathrm{l},0} + \beta_{Q_\mathrm{a,l},1} \left(Q_\mathrm{a}\right) + \beta_{E_\mathrm{ch,l},1} \left(E_\mathrm{ch}\right),
\end{equation}
where $\beta_{\mathrm{l},0}$ represents the intercept of the linear regression, $\beta_{Q_\mathrm{a,l},1}$ the linear dependency on the annual battery production, and $\beta_{E_\mathrm{ch,l},1}$ the linear dependency on the carbon intensity of the Chinese electricity mix.
Second, we apply a logarithmic transformation to frame the problem as a linear or multi-linear regression, a common technique for estimating coefficients in power regressions~\citep{MaharjanHauckEtAl2024,LouwenJunginger2021}, obtaining 
\begin{equation}
	\log (P_{\mathrm{gw}})  = \beta_{\mathrm{p},0} + \beta_{Q_\mathrm{a,p},1} \log \left(Q_\mathrm{a}\right),
\end{equation}
\begin{equation}
	\log(P_{\mathrm{gw}})  = \beta_{\mathrm{p},0} + \beta_{E_\mathrm{ch,p},1} \log \left(E_\mathrm{ch}\right),
\end{equation}
\begin{equation}
	\log(P_{\mathrm{gw}})  = \beta_{\mathrm{p},0} + \beta_{Q_\mathrm{a,p},1} \log \left(Q_\mathrm{a}\right) + \beta_{E_\mathrm{ch,p},1} \log \left(E_\mathrm{ch}\right),
\end{equation}
with $\beta_{\mathrm{p},0}$ representing the constant terms, $\beta_{Q_\mathrm{a,p},1}$ the logarithmic coefficient for the annual battery production, and $\beta_{E_\mathrm{ch,p},1}$ the logarithmic coefficient for the carbon intensity of the Chinese electricity mix.
We identify the coefficients by solving an error minimization problem between the data points $\hat{GWP}_i$ and our predicted value. For Eq.~\eqref{eq:GWP-Q-l}, the problem takes the form
\begin{equation}
	\min_{\beta_{\mathrm{l},0} \: \beta_{Q_\mathrm{a,l},1}} \sum_{i=1}^{N} \left(\hat{P}_{\mathrm{gw},i} - \beta_{\mathrm{l},0} - \beta_{Q_\mathrm{a,l},1} \left(Q_\mathrm{a}\right) \right)^2,
\end{equation}
and can be framed for the other regressions in a similar fashion.
Ultimately, we present six models to estimate $P_{\mathrm{gw}}$ and determine which one best matches the data.

\begin{Figure}
	\centering
	\includegraphics[width=\columnwidth]{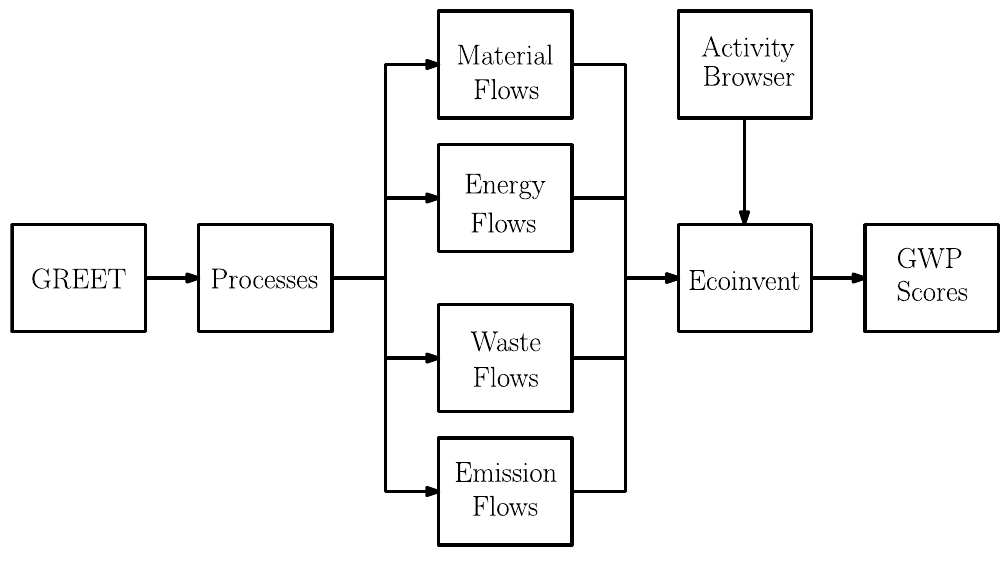}
	\captionof{figure}{Flow diagram of the cradle-to-gate \ac{LCA}. We use the GREET battery inventory to quantify the flows and we retrieve the associated GWP scores from Ecoinvent, using Activity Browser to change parameters such as regional electricity mix.}
	\label{fig:LCA1}
\end{Figure}

\begin{Figure}
	\centering
	\includegraphics[width=0.95\columnwidth]{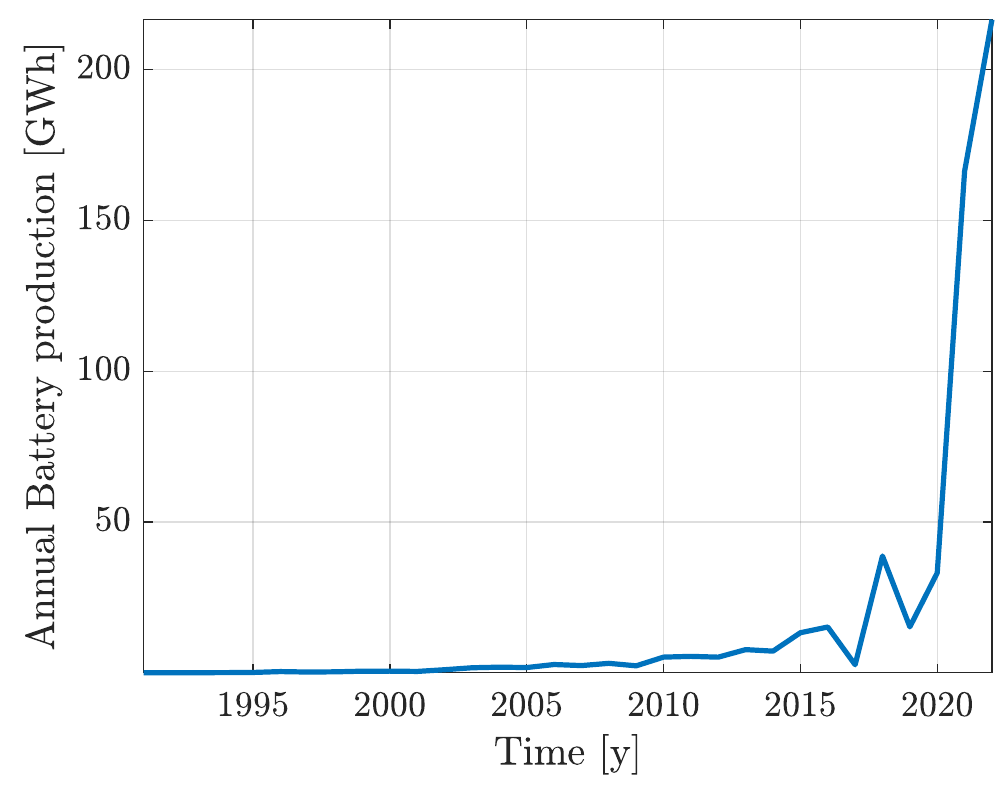}
	\captionof{figure}{Annual battery production volumes~\citep{IEA2021,Placek2021}.}
	\label{fig:trend_AN}
\end{Figure}

\begin{Figure}
	\centering
	\includegraphics[width=\columnwidth]{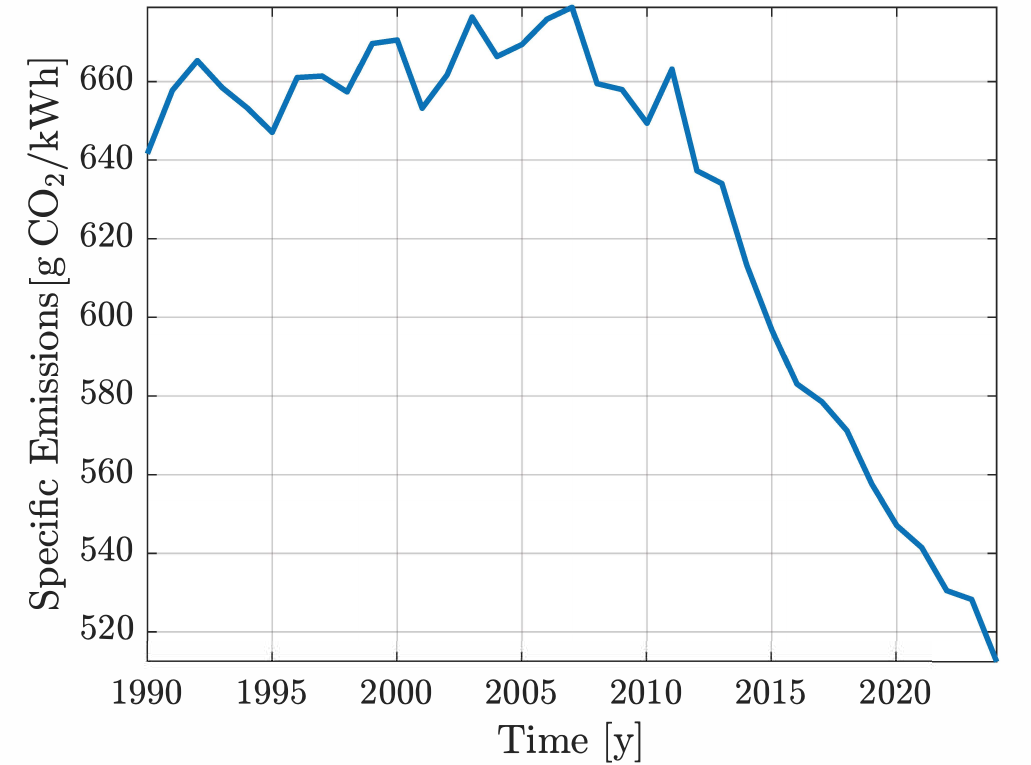}
	\captionof{figure}{Carbon intensity of the Chinese electricity mix~\citep{LowCarbonPower2024}.}
	\label{fig:trend_CH}
\end{Figure}



\subsection{Discussion}\label{subsec:disc}

In this paragraph, we present some clarification on the assumptions and limitations of our work.
First, we assume a standard production process to evaluate contributions, consistent with the fact that in China the three largest producers (CATL, BYD and Gotion) account for nearly 50\% of the country’s production capacity~\citep{IEA2024}.
Second, we consider the sources of our meta-analysis and regression models independent of each other. However, some studies rely on secondary data from literature or databases like Ecoinvent, GREET, and GaBi.
Finally, in the forecasts, we assume no spillover learning effects from other technologies, considering only the impact of our predictors.


\section{Results} \label{sec:Results2}

In this section, we present the numerical results of the study.
Section~\ref{subsec:LCAO} details the findings of the \ac{LCA}, demonstrating the flexibility of our framework through a case study on variable electricity mixes and comparing the results with the existing literature.
Lastly, Section~\ref{subsec:RM} discusses the statistical analysis of the regression models.

\subsection{Life-cycle Assessment Outcome}~\label{subsec:LCAO}

\begin{table*}[b]
	\centering
	\caption{Summary of regression results and model selection criteria. The Akaike (AIC) and Bayesian (BIC) information criteria assess the trade-off between goodness-of-fit and model complexity.}
	\label{tab:RES}
	\resizebox{\textwidth}{!}{
	\begin{tabular}{l | l| c c c c | c c c c}
		\toprule
		\multirow{2}{*}{\textbf{Regression}} & \multirow{2}{*}{\textbf{Predictor(s)}} & \multicolumn{4}{c}{\textbf{Original Data (without outliers) N = 36}} & \multicolumn{4}{c}{\textbf{Averaged Data (without outliers) N = 7 }} \\
																		&				&  \textbf{R$^2$ (adj-R$^2$)} & \textbf{p-value} & \textbf{AIC} & \textbf{BIC} & \textbf{R$^2$ (adj-R$^2$)} & \textbf{p-value} & \textbf{AIC} & \textbf{BIC}\\
		\midrule

		\multirow{3}{*}{\textbf{Linear}} & $Q_\mathrm{a}$ & 0.0028 (-0.0265) & 0.7585 & 248.5186 &251.6856 &0.0610 (-0.1268) & 0.5935 & 37.9395 & 37.8313 \\
											& $E_\mathrm{ch}$ &0.0021 (-0.0272) &0.7900 & 248.5440 & 251.7110 & 0.0206 (-0.1753) &0.7589 &38.2342 &38.1261 \\
											&$Q_\mathrm{a}$, $E_\mathrm{ch}$ & 0.0034 (-0.0570) & 0.9457 & 250.4984 & 255.2489 & 0.6034 (0.4051) & 0.1573 &33.9057 & 33.7434 \\
				\midrule
		\multirow{3}{*}{\textbf{Log-Log}} & 	$Q_\mathrm{a}$ & 0.0092 (-0.0199) & 0.5778 & 40.7412 &43.9082 & 0.0374 (-0.1551) & 0.6777 & -6.0955 & -6.2037 \\
													&	$E_\mathrm{ch}$ & 0.0004 (-0.0290) & 0.9130& 41.0613 & 44.2283 & 0.0197 (-0.1764) & 0.7641 & -5.9678 & -6.0760 \\
													&	$Q_\mathrm{a}$, $E_\mathrm{ch}$ & 0.0098 (-0.0502) & 0.8499 & 42.7192 & 47.4697 & 0.0479 (-0.4281) &0.9065 &-4.1722 &-4.3345 \\

		\bottomrule
	\end{tabular}
	}
\end{table*}

In our \ac{LCA} study, we stage a large part of the production process in China, as it is home to more than three-quarters of the installed NMC production capacity~\citep{IEA2024}. 
Even in case studies involving products assembled in third countries, most of the materials are primarily processed in China, as evidenced by data showing that approximately 80\% of all cobalt products are used for battery production in China~\citep{USGS2022}.
However, we still include the contribution of other processes and materials, such as Aluminum and the Battery Management System, based on global average values.

\begin{figure*}[ht!]
	\centering
	\includegraphics[width=\linewidth]{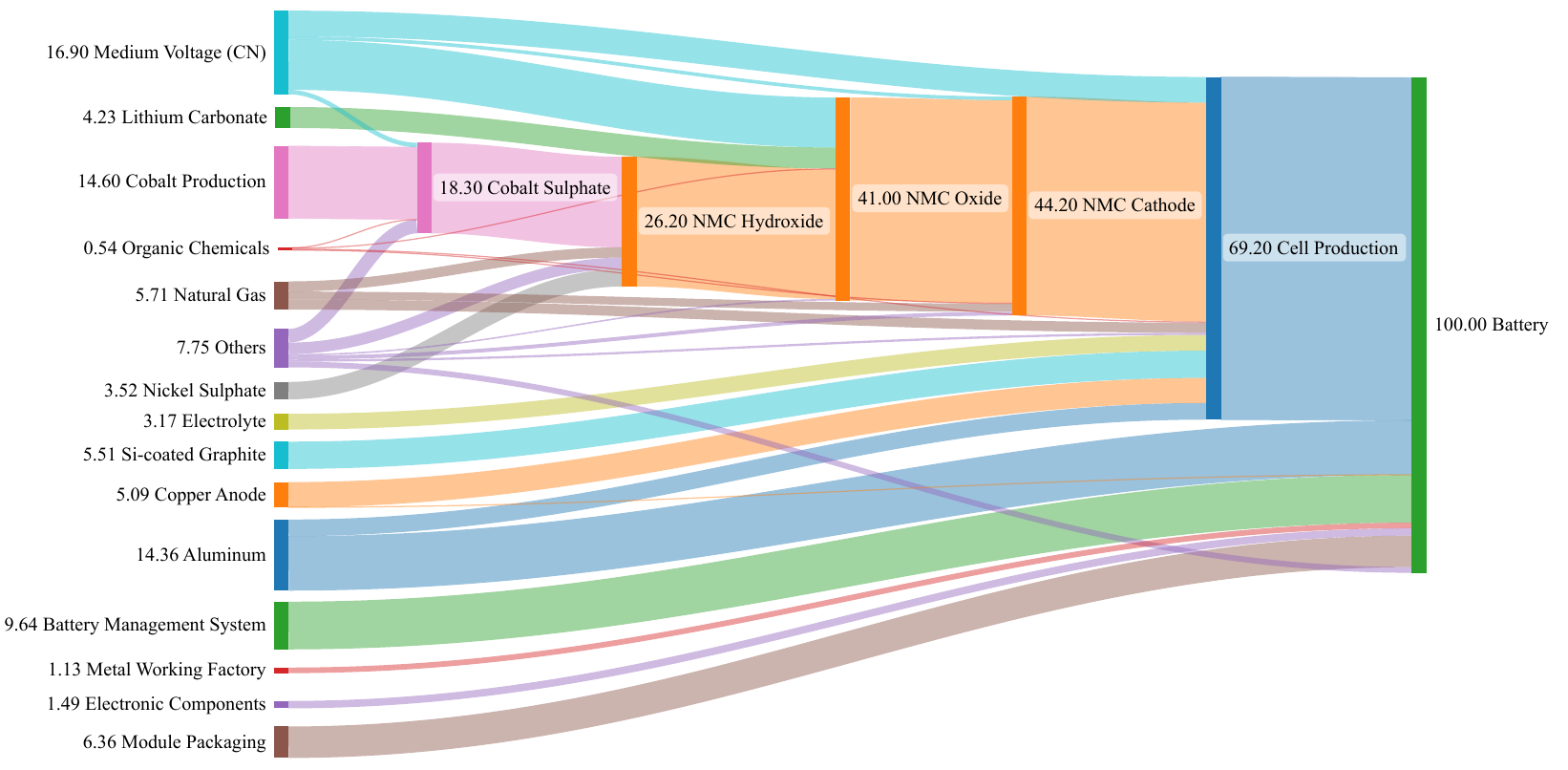}
	\caption{Percentual composition of the \acp{GHG} considering the Chinese electricity mix.}
	\label{fig:sankey}
\end{figure*}

\begin{Table}
	\centering	
	\captionof{table}{GWP values for China, South Korea, and Sweden}
	\resizebox{\linewidth}{!}{
	\begin{tabular}{l cc}
		\toprule
		\textbf{Country} & \multicolumn{2}{c}{\textbf{Global Warming Potential}} \\
		\cmidrule(lr){2-3}
		& \unit{kg} CO\textsubscript{2}-eq.\ per \unit{kg} & \unit{kg} CO\textsubscript{2}-eq.\ per \unit{kWh} \\
		\midrule
		China        & 17.33 & 82.92 \\
		South Korea  & 16.85 & 80.62 \\
		Sweden       & 16.47 & 78.80 \\
		\bottomrule
	\end{tabular}
	}
	\label{tab:GWP_values}
\end{Table}

\begin{figure*}[ht!]
	\centering
	\includegraphics[width=\linewidth]{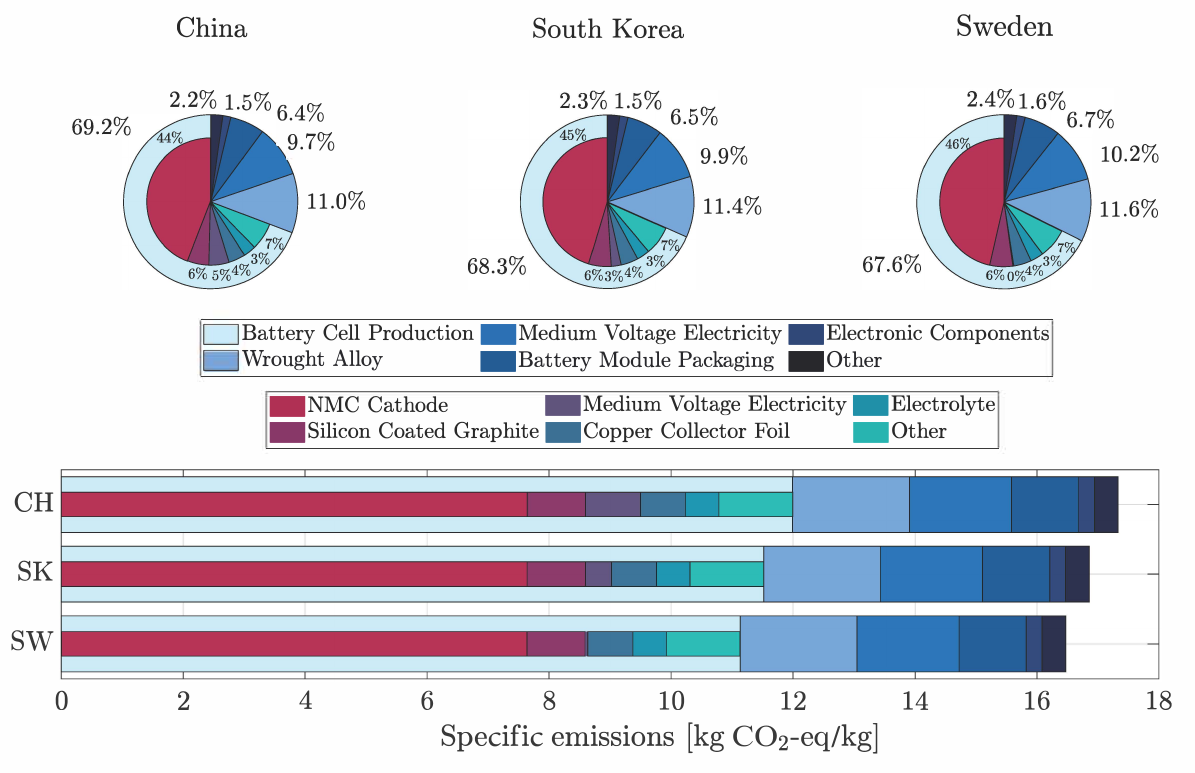}
	\caption{Influence of the electricity mix on the \ac{LCA} of mass-specific \acp{GHG} emissions in the production of Li-Ion batteries. We consider the electricity mixes from South Korea (SK), Sweden (SE), and the global average (GLO).}
	\label{fig:emissions}
\end{figure*}

We showcase the flexibility of the \ac{LCA} framework by analyzing the manufacturing process in countries with varying levels of carbon intensity of electricity mix: China, Sweden, and South Korea.
We compute the \ac{GWP} scores using the approach described in Section~\ref{subsec:LCA}, detailing the several contribution and their impact in Appendix~\ref{app:LCAA} in Tables \ref{tab:BTCH}, \ref{tab:BTSK}, and \ref{tab:BTSW} for 1 \unit{kg} of battery, and in Tables \ref{tab:CELLCH}, \ref{tab:CELLSK}, and \ref{tab:CELLSW} for 1 \unit{kg} of cell. 
In Fig.~\ref{fig:sankey}, we analyze the decomposition of these contributions to the overall GWP considering a Sankey diagram for the Chinese case study, while in Fig.~\ref{fig:emissions} we compare the impact of each source for the different countries.
We observe that the major contribution is cell production, and at a deeper level, the energy used to process the cathode materials, which aligns with the findings of the meta-analysis.
When comparing the major contributions across different case studies, we find very similar results, differing significantly only with respect to the impact of the medium voltage electricity used in the cell production.
This finding indicates that, unless the entire production line is relocated, assembling the battery in a country with a clean energy mix may have minimal effect on the environmental impact.
Finally, Table~\ref{tab:GWP_values} presents the specific \ac{GWP} of Li-Ion \ac{NMC} batteries in \unit{kg} CO$_2$-eq. per \unit{kg} and a conversion to \unit{kg} CO$_2$-eq. per \unit{kWh} using a specific energy capacity of 0.209 \unit{kWh/kg}.
When comparing these results with the meta-analysis in Section~\ref{subsec:Meta}, they fall perfectly within the statistical range, closely aligning with the median of the set without outliers.


%

\subsection{Statistical Analysis}~\label{subsec:RM}


Of all the proposed regression models in Table~\ref{tab:RES}, only one of the linear regressions for the yearly-averaged dataset shows statistical significance, specifically, the one using annual production and carbon intensity of the Chinese electricity mix as predictors. 
A statistical analysis of the regression indicates a moderate explanatory power of the predictors even with a small number of data points, with an R$^2$ value of 0.6034. 
Moreover, the overall p-value of 0.1573 suggests that this regression model fits the data better than a model without independent variables, using a significance level ($\alpha$) lower than what is typically applied in conventional \ac{LCA} studies.
The positive coefficient of $\beta_{\mathrm{1,e}}$  links a reduction in carbon intensity to a lower specific GWP, as also suggested by the meta-analysis and our case study. 
Conversely, the negative coefficient for production volume indicates that an increase in production is linked to a lower specific GWP, backing the premise of learning effects taking place.
Furthermore, the analysis of the residuals finds no significant auto-correlation due to a value of 2.2035 of the Durbin-Watson Statistic. 
The normal distribution of the residuals, testified by the values of the parameters in Table~\ref{tab:residuals_analysis}, wards off any concerns about heteroskedasticity.
Ultimately, the assumptions of normality and independence are reasonably satisfied, although the small number of data points suggests that this explanatory power should be interpreted with caution.

\begin{Table}
	\centering
	\captionof{table}{Regression Coefficients for the statistically significant regression model.}
	\label{tab:regression_coefficients}
		\resizebox{\linewidth}{!}{
		\begin{tabular}{ l c c c c}
			\toprule
			\textbf{Coefficient} & \textbf{Value} & \textbf{$\sigma$} & \textbf{t-Statistic} & \textbf{p-value} \\
			\midrule
			$\beta_\mathrm{0}$ & -185.7 & 89.266 & -2.0803 & 0.10599 \\
			$\beta_{\mathrm{1,a}}$ & -1.2162 & 0.50161 & -3.4246 & 0.072404 \\
			$\beta_{\mathrm{1,e}}$ & 0.38658 & 0.16527 & 2.3391 & 0.079458 \\
			\bottomrule
		\end{tabular}
		}
\end{Table}

\begin{Table}
	\centering
	\captionof{table}{Residuals Analysis Summary}
	\label{tab:residuals_analysis}
		\resizebox{\linewidth}{!}{
		\begin{tabular}{l c c c c}
			\toprule
			& \textbf{Durbin-Watson} & \textbf{Skewness} & \textbf{Kurtosis} & $\boldsymbol{\chi^2}$ \\
			\midrule
			\textbf{Value} & 2.2035 & -0.5303 & 2.5922 & 0.3766 \\
			\bottomrule
		\end{tabular}
		}
\end{Table}

\section{Conclusion} \label{sec:Conclusions}

This study examined how annual production volume and electricity mix influence the mass-specific global warming potential of rechargeable, prismatic \ac{NMC} 811 lithium-ion batteries with silicon-coated graphite anodes. 
Through a comprehensive meta-analysis of the past two decades of \ac{LCA} studies, we compiled a dataset to benchmark our results and assess trends in environmental impact.
The analysis identified a median global warming potential of 17.63~\unit{kg} CO$_2$-eq./\unit{kg} with a standard deviation of 7.34~\unit{kg} CO$_2$-eq./\unit{kg}, across published studies.
Our cradle-to-gate assessment, applied across different national electricity mixes, yielded values consistent with the statistical median: 17.33~\unit{kg} CO$_2$-eq./\unit{kg} for China, 16.85~\unit{kg} CO$_2$-eq./\unit{kg} for South Korea, and 16.47~\unit{kg} CO$_2$-eq./\unit{kg} for Sweden.
The relatively small differences across these cases indicate that upstream carbon-intensive processes, especially cathode material production in China, limit the potential benefits of relocating final assembly to regions with cleaner electricity mixes. Achieving meaningful reductions in the carbon footprint requires decarbonization efforts across the entire battery supply chain.
The regression analysis links production scale and the carbon intensity of the Chinese energy mix to specific emissions, supporting the presence of learning effects in battery manufacturing. 
A linear model incorporating both factors achieved moderate explanatory power (R$^2$ = 0.6034), performing significantly better than a baseline model without independent variables. However, given the limited number of data points, this conclusion should be interpreted with caution.

This work opens the field for the following extensions: 
First, we aim to collect more data points to enhance the model's accuracy and reliability.
Second, we would like to include a learn-by-searching effect by incorporating the number of patents in the predictors and investigate further correlations.
Finally, we intend to carry out a multicollinearity analysis of the independent variables considered.

\begin{acknowledgements}
We thank Dr.~I.~New for proofreading this paper.
This publication is part of the project \cite{NEON} (with project number 17628 of the research program Crossover which is (partly) financed by the Dutch Research Council (NWO)).
\end{acknowledgements}

\begin{small}

\bibliography{main.bib, JCP.bib, SML_papers.bib,MCL.bib, RCR.bib,references.bib}

\bibliographystyle{spbasic}      



\end{small}

\begin{appendices}
	\appendix
\section{Battery Life-cycle Assessment} \label{app:LCAA}

The inventories, amounts (in \unit{kg}), and the \ac{LCA} score for the different case studies is reported, respectively for China, South Korea, and Sweden, in Tables \ref{tab:BTCH}, \ref{tab:BTSK}, and \ref{tab:BTSW} for a \unit{kg} of battery, and in Tables \ref{tab:CELLCH}, \ref{tab:CELLSK}, and \ref{tab:CELLSW} for a \unit{kg} of cell.
The total score of the cell differs from the first entry of the battery as the amount required is lower than 1~\unit{kg}. Dividing the score by the amount in the Tables referring to the battery we find exactly the total score for the cell.

\begin{table*}[b]
	\centering
	\caption{Battery Life Cycle Assessment for China\\}
	\resizebox{\textwidth}{!}{
		\begin{tabular}{l l l l}
			\toprule
			\textbf{Exchange Input} & \textbf{Origin} & \textbf{Amount (kg)} & \textbf{Score} \\
			\midrule
			Mrk. for battery cell production, Li-ion, NMC811 & CN & 0.71359 & 11.99286 \\
			Mrk. for aluminium, wrought alloy & GLO & 0.14283 & 1.91437 \\
			Mrk. for battery management system production, Li-ion & GLO & 0.02426 & 1.67246 \\
			Mrk. for battery module packaging, Li-ion & GLO & 0.05718 & 1.10324 \\
			Mrk. for electronic component, passive, unspecified & GLO & 0.00431 & 0.25795 \\
			Mrk. for metal working factory & GLO & 1.48e-09 & 0.18149 \\
			Mrk. for impact extrusion of aluminium, 1 stroke & GLO & 0.14162 & 0.12155 \\
			Mrk. for ethylene glycol & GLO & 0.02302 & 0.04556 \\
			Mrk. for reinforcing steel & GLO & 0.00642 & 0.01323 \\
			Mrk. for polyethylene, high density, granulate & GLO & 0.00405 & 0.00909 \\
			Mrk. for copper, anode & GLO & 0.00100 & 0.00576 \\
			Mrk. for injection moulding & GLO & 0.00405 & 0.00499 \\
			Mrk. for glass fibre reinforced plastic, polyamide, injection moulded & GLO & 0.00033 & 0.00288 \\
			Mrk. for sheet rolling, steel & GLO & 0.00642 & 0.00230 \\
			Mrk. for sheet rolling, aluminium & GLO & 0.00121 & 0.00077 \\
			Mrk. for sheet rolling, copper & GLO & 0.00100 & 0.00052 \\
			Mrk. group for electricity, medium voltage & CN & 0.00028 & 2.79e-04 \\
			Mrk. for tap water & RoW & 0.02302 & 2.34e-05 \\
			\midrule
			& \textbf{Total} & & 17.3293 \\
			\bottomrule
		\end{tabular}
	}
	\label{tab:BTCH}
	
\end{table*}

\begin{table*}
	\centering
	\caption{Battery Life Cycle Assessment for South Korea\\}
	\resizebox{\textwidth}{!}{
		\begin{tabular}{l l l l}
			\toprule
			\textbf{Exchange Input} & \textbf{Origin} & \textbf{Amount (kg)} & \textbf{Score} \\
			\midrule
			Mrk. for battery cell production, Li-ion, NMC811 & SK & 0.71359 & 11.51889 \\
			Mrk. for aluminium, wrought alloy & GLO & 0.14283 & 1.91437 \\
			Mrk. for battery management system production, Li-ion & GLO & 0.02426 & 1.67246 \\
			Mrk. for battery module packaging, Li-ion & GLO & 0.05718 & 1.10324 \\
			Mrk. for electronic component, passive, unspecified & GLO & 0.00431 & 0.25795 \\
			Mrk. for metal working factory & GLO & 1.48e-09 & 0.18149 \\
			Mrk. for impact extrusion of aluminium, 1 stroke & GLO & 0.14162 & 0.12155 \\
			Mrk. for ethylene glycol & GLO & 0.02302 & 0.04556 \\
			Mrk. for reinforcing steel & GLO & 0.00642 & 0.01323 \\
			Mrk. for polyethylene, high density, granulate & GLO & 0.00405 & 0.00909 \\
			Mrk. for copper, anode & GLO & 0.00100 & 0.00576 \\
			Mrk. for injection moulding & GLO & 0.00405 & 0.00499 \\
			Mrk. for glass fibre reinforced plastic, polyamide, injection moulded & GLO & 0.00033 & 0.00288 \\
			Mrk. for sheet rolling, steel & GLO & 0.00642 & 0.00230 \\
			Mrk. for sheet rolling, aluminium & GLO & 0.00121 & 0.00077 \\
			Mrk. for sheet rolling, copper & GLO & 0.00100 & 0.00052 \\
			Mrk. for electricity, medium voltage & SK & 0.00028 & 1.32e-04 \\
			Mrk. for tap water & RoW & 0.02302 & 2.34e-05 \\
			\midrule
			& \textbf{Total} & & 16.8552 \\
			\bottomrule
		\end{tabular}
	}
	\label{tab:BTSK}
	
\end{table*}

\begin{table*}
	\centering
	\caption{Battery Life Cycle Assessment for Sweden\\}
	\resizebox{\textwidth}{!}{
		\begin{tabular}{l l l l}
			\toprule
			\textbf{Exchange Input} & \textbf{Origin} & \textbf{Amount (kg)} & \textbf{Score} \\
			\midrule
			Mrk. for battery cell production, Li-ion, NMC811 & SE & 0.71359 & 11.13418 \\
			Mrk. for aluminium, wrought alloy & GLO & 0.14283 & 1.91437 \\
			Mrk. for battery management system production, Li-ion & GLO & 0.02426 & 1.67246 \\
			Mrk. for battery module packaging, Li-ion & GLO & 0.05718 & 1.10324 \\
			Mrk. for electronic component, passive, unspecified & GLO & 0.00431 & 0.25795 \\
			Mrk. for metal working factory & GLO & 1.48e-09 & 0.18149 \\
			Mrk. for impact extrusion of aluminium, 1 stroke & GLO & 0.14162 & 0.12155 \\
			Mrk. for ethylene glycol & GLO & 0.02302 & 0.04556 \\
			Mrk. for reinforcing steel & GLO & 0.00642 & 0.01323 \\
			Mrk. for polyethylene, high density, granulate & GLO & 0.00405 & 0.00909 \\
			Mrk. for copper, anode & GLO & 0.00100 & 0.00576 \\
			Mrk. for injection moulding & GLO & 0.00405 & 0.00499 \\
			Mrk. for glass fibre reinforced plastic, polyamide, injection moulded & GLO & 0.00033 & 0.00288 \\
			Mrk. for sheet rolling, steel & GLO & 0.00642 & 0.00230 \\
			Mrk. for sheet rolling, aluminium & GLO & 0.00121 & 0.00077 \\
			Mrk. for sheet rolling, copper & GLO & 0.00100 & 0.00052 \\
			Mrk. for tap water & RoW & 0.02302 & 2.34e-05 \\
			Mrk. for electricity, medium voltage & SE & 0.00028 & 1.22e-05 \\
			\midrule
			& \textbf{Total} & & 16.4704 \\
			\bottomrule
		\end{tabular}
	}
	\label{tab:BTSW}
	
\end{table*}

\begin{table*}
	\centering
	\caption{Battery Cell's Life Cycle Assessment for China\\}
	\resizebox{\textwidth}{!}{
		\begin{tabular}{l l l l}
			\toprule
			\textbf{Exchange Input} & \textbf{Origin} & \textbf{Amount (kg)} & \textbf{Score} \\
			\midrule
			Mrk. for cathode, NMC811, for Li-ion battery & CN & 0.37730 & 10.70836 \\
			Mrk. for anode, silicon coated graphite, for Li-ion battery & CN & 0.21810 & 1.34061 \\
			Mrk. group for electricity, medium voltage & CN & 1.26160 & 1.25836 \\
			Mrk. for copper collector foil, for Li-ion battery & GLO & 0.12430 & 1.03297 \\
			Mrk. for electrolyte, for Li-ion battery & GLO & 0.16830 & 0.77075 \\
			Mrk. for heat, district or industrial, natural gas & RoW & 13.29100 & 0.49437 \\
			Mrk. for aluminium collector foil, for Li-ion battery & GLO & 0.02840 & 0.43461 \\
			Mrk. for aluminium, wrought alloy & GLO & 0.02840 & 0.38065 \\
			Mrk. for copper, anode & GLO & 0.03300 & 0.18994 \\
			Mrk. for battery separator & GLO & 0.01820 & 0.08900 \\
			Mrk. for chemical factory, organics & GLO & 4e-10 & 0.05841 \\
			Mrk. for sheet rolling, aluminium & GLO & 0.02840 & 0.01797 \\
			Mrk. for sheet rolling, copper & GLO & 0.03300 & 0.01708 \\
			Mrk. for polyethylene terephthalate, granulate, amorphous & GLO & 0.00280 & 0.00851 \\
			Mrk. for polypropylene, granulate & GLO & 0.00120 & 0.00266 \\
			Mrk. for extrusion, plastic film & GLO & 0.00400 & 0.00213 \\
			\midrule
			& \textbf{Total} & & 16.8064 \\
			\bottomrule
		\end{tabular}
	}
	\label{tab:CELLCH}
	
\end{table*}

\begin{table*}
	\centering
	\caption{Battery Cell's Life Cycle Assessment for South Korea\\}
	\resizebox{\textwidth}{!}{
		\begin{tabular}{l l l l}
			\toprule
			\textbf{Exchange Input} & \textbf{Origin} & \textbf{Amount (kg)} & \textbf{Score} \\
			\midrule
			Mrk. for cathode, NMC811, for Li-ion battery & CN & 0.37730 & 10.70836 \\
			Mrk. for anode, silicon coated graphite, for Li-ion battery & CN & 0.21810 & 1.34061 \\
			Mrk. for copper collector foil, for Li-ion battery & GLO & 0.12430 & 1.03297 \\
			Mrk. for electrolyte, for Li-ion battery & GLO & 0.16830 & 0.77075 \\
			Mrk. for electricity, medium voltage & SK & 1.26160 & 0.59415 \\
			Mrk. for heat, district or industrial, natural gas & RoW & 13.29100 & 0.49437 \\
			Mrk. for aluminium collector foil, for Li-ion battery & GLO & 0.02840 & 0.43461 \\
			Mrk. for aluminium, wrought alloy & GLO & 0.02840 & 0.38065 \\
			Mrk. for copper, anode & GLO & 0.03300 & 0.18994 \\
			Mrk. for battery separator & GLO & 0.01820 & 0.08900 \\
			Mrk. for chemical factory, organics & GLO & 4e-10 & 0.05841 \\
			Mrk. for sheet rolling, aluminium & GLO & 0.02840 & 0.01797 \\
			Mrk. for sheet rolling, copper & GLO & 0.03300 & 0.01708 \\
			Mrk. for polyethylene terephthalate, granulate, amorphous & GLO & 0.00280 & 0.00851 \\
			Mrk. for polypropylene, granulate & GLO & 0.00120 & 0.00266 \\
			Mrk. for extrusion, plastic film & GLO & 0.00400 & 0.00213 \\
			\midrule
			& \textbf{Total} & & 16.1422 \\
			\bottomrule
		\end{tabular}
	}
	\label{tab:CELLSK}
	
\end{table*}

\begin{table*}
	\centering
	\caption{Battery Cell's Life Cycle Assessment for Sweden\\}
	\resizebox{\textwidth}{!}{
		\begin{tabular}{l l l l}
			\toprule
			\textbf{Exchange Input} & \textbf{Origin} & \textbf{Amount (kg)} & \textbf{Score} \\
			\midrule
			Mrk. for cathode, NMC811, for Li-ion battery & CN & 0.37730 & 10.70836 \\
			Mrk. for anode, silicon coated graphite, for Li-ion battery & CN & 0.21810 & 1.34061 \\
			Mrk. for copper collector foil, for Li-ion battery & GLO & 0.12430 & 1.03297 \\
			Mrk. for electrolyte, for Li-ion battery & GLO & 0.16830 & 0.77075 \\
			Mrk. for heat, district or industrial, natural gas & RoW & 13.29100 & 0.49437 \\
			Mrk. for aluminium collector foil, for Li-ion battery & GLO & 0.02840 & 0.43461 \\
			Mrk. for aluminium, wrought alloy & GLO & 0.02840 & 0.38065 \\
			Mrk. for copper, anode & GLO & 0.03300 & 0.18994 \\
			Mrk. for battery separator & GLO & 0.01820 & 0.08900 \\
			Mrk. for chemical factory, organics & GLO & 4e-10 & 0.05841 \\
			Mrk. for electricity, medium voltage & SE & 1.26160 & 0.05504 \\
			Mrk. for sheet rolling, aluminium & GLO & 0.02840 & 0.01797 \\
			Mrk. for sheet rolling, copper & GLO & 0.03300 & 0.01708 \\
			Mrk. for polyethylene terephthalate, granulate, amorphous & GLO & 0.00280 & 0.00851 \\
			Mrk. for polypropylene, granulate & GLO & 0.00120 & 0.00266 \\
			Mrk. for extrusion, plastic film & GLO & 0.00400 & 0.00213 \\
			\midrule
			& \textbf{Total} & & 15.6031 \\
			\bottomrule
		\end{tabular}
	}
	\label{tab:CELLSW}
	
\end{table*}
\end{appendices}

\end{multicols}
\end{document}